\documentclass[journal]{IEEEtran}
\usepackage{amsmath, amssymb}
\usepackage{graphicx}
\usepackage{cite}
\usepackage{url}
\usepackage{caption}   
\usepackage{booktabs}  
\usepackage{placeins}
\usepackage[hidelinks]{hyperref}

\title{MusRec: Zero-Shot Text-to-Music Editing via Rectified Flow and Diffusion Transformers}
\author{Ali Boudaghi,
        Hadi Zare%
\thanks{Corresponding author: Ali Boudaghi (email: ali.boudaghi@ut.ac.ir).}}

\begin{document}

\maketitle

\begin{abstract}
Music editing has emerged as an important and practical area of artificial intelligence, with applications ranging from video game and film music production to personalizing existing tracks according to user preferences. However, existing models face significant limitations, such as being restricted to editing synthesized music generated by their own models, requiring highly precise prompts, or necessitating task-specific retraining—thus lacking true zero-shot capability. leveraging recent advances in rectified flow and diffusion transformers, we introduce MusRec, a zero-shot text-to-music editing model capable of performing diverse editing tasks on real-world music efficiently and effectively. Experimental results demonstrate that our approach outperforms existing methods in preserving musical content, structural consistency, and editing fidelity, establishing a strong foundation for controllable music editing in real-world scenarios.
\end{abstract}

\begin{IEEEkeywords}
Music Editing, Diffusion Models, Rectified Flow, Audio Generation, Zero-Shot Learning.
\end{IEEEkeywords}

\section{Introduction}

The landscape of audio generation has shifted dramatically in recent years. 
Text-to-music systems now allow users to compose entire musical pieces from simple textual descriptions, powered by advances in diffusion models and transformer architectures~\cite{forsgren2022riffusion, liu2023audioldm, huang2023noise2music, Aufussion, accomontage, musicldm,lin2023content, musecoco, copet2024simple,melechovsky2023mustango, textcond}. 
While impressive, these systems are still primarily designed for \emph{creation from scratch}. 
In contrast, real-world music practice often revolves around \emph{editing}: refining a performance, altering instrumentation, or adapting an existing recording into a new style. 
For musicians, producers, and casual creators alike, the ability to reshape existing audio is often more valuable than generating entirely new material.

Music editing, however, is fundamentally more difficult than generation. 
It requires the model to balance two competing goals: applying the requested modification faithfully, and preserving the rich details of the input recording that should remain unchanged. 
This trade-off is especially challenging when dealing with expressive, polyphonic, or multi-instrumental recordings. 
Existing research has attempted to address editing through supervised datasets of paired ``before'' and ``after'' examples~\cite{m2ugen, instructme, audit}, or through zero-shot latent manipulations in diffusion models~\cite{musicmagus, zeta, transplayer}. 
Yet most methods remain restricted by their limitation to specific editing tasks, operate mainly on model-generated music rather than arbitrary recordings, and often require very precise prompts to succeed~\cite{musicmagus, transplayer}. 
These limitations hinder their use in flexible, user-friendly creative workflows. Recent works also show that diffusion models can be effective for audio restoration tasks, such as equalization and bandwidth extension~\cite{moliner2024diffusion_equalizer}.

At the same time, a parallel line of research has introduced rectified flow models~\cite{lee2024improvingtrainingrectifiedflows, liu2022flowstraightfastlearning}, which reformulate diffusion as a more direct flow between noise and data distributions. 
Rectified flows enable efficient and stable generation, and have recently been realized at scale through the \emph{Flux} family of diffusion transformers~\cite{tang2024hart, xie2024sana, yang2024cogvideox, peebles2023scalable}. 
\emph{FLUX that Plays Music}~\cite{fluxplaysmusic} in particular demonstrated the power of this approach for text-to-music generation. 
In computer vision, the work of \emph{Taming Rectified Flow for Inversion and Editing}~\cite{tamingrectifiedflow} showed that RF models also support accurate inversion and robust editing, but these ideas have not yet been applied to music. 
This raises an intriguing question: can the strengths of rectified flow be used not just for generating music, but for editing real recordings in a practical, zero-shot fashion?

\subsection{Our Approach}

In this paper, we introduce a framework for zero-shot music editing based on rectified flow models.
Our approach is motivated by recent progress in improving rectified flow (RF) inversion and editing. 
RF-Solver~\cite{tamingrectifiedflow} addresses the reconstruction problem by formulating a more precise sampler for solving the RF ODE, reducing error accumulation during inversion and thereby yielding more faithful reconstructions. 
Building on this, RF-Edit~\cite{tamingrectifiedflow} extends the idea to practical image and video editing: it stabilizes edits by storing and re-injecting the V (value) feature in the
self-attention layers of the source, which preserves structure while allowing targeted modifications. 
Inspired by these advances, we adapt the principles of RF-Solver and RF-Edit to the audio domain.

Specifically, we leverage a Flux-style diffusion transformer originally trained for text-to-music generation~\cite{fluxplaysmusic}, and extend its capability to real-audio editing through an inversion procedure that maps raw recordings into the rectified-flow latent space. 
Within this space, targeted manipulations---such as timbre transfer between instruments---can be performed before decoding the results back into high-fidelity music audio.

Our design deliberately avoids additional training: the entire pipeline works in a \emph{zero-shot} setting. 
This choice offers several concrete advantages over prior editing approaches:
\begin{enumerate}
    \item \textbf{Zero-shot editing}: no fine-tuning, paired data, or supervision is required. 
    \item \textbf{Real-audio compatibility}: the method accepts arbitrary recordings as inputs, not just outputs generated by the model itself. 
    \item \textbf{Instrument-agnostic timbre transfer}: edits are not tied to a fixed instrument vocabulary, allowing flexible cross-instrument transformations. 
    \item \textbf{Accessible prompting}: coarse or natural descriptions suffice, removing the need for carefully engineered text prompts. 
    \item \textbf{Efficient inversion and generation}: our method performs both inversion and editing in only 25 diffusion steps, whereas other models typically require between 50 and 200 steps to achieve comparable results.
\end{enumerate}

\subsection{Contributions}

In summary, our work introduces a new perspective on music editing:
\begin{itemize}
    \item We present rectified-flow based framework for editing of real music recordings. 
    \item We demonstrate a zero-shot pipeline that performs timbre transfer, genre transfer and other edits without any re-training. 
    \item We highlight the practical advantages of this approach: generality across instruments and genres, compatibility with real audio, user-friendly interaction, and fast inversion and generation. 
    \item Through experiments on diverse datasets and metrics, we show that our method maintains fidelity to the input recording while applying edits with high transferability. 
\end{itemize}

By extending rectified flow beyond generation into editing, we reveal its potential as a foundation for flexible, high-quality, and accessible tools for music creation and transformation.

\section{Related Work}

\subsection{Text-to-Music Generation} 
Text-to-music generation has seen rapid advances with the rise of diffusion and transformer-based models. Early approaches relied on autoregressive language models applied to audio data~\cite{agostinelli2023musiclm, copet2024simple, musicgenstem, inspiremusic}. Autoregressive models are advantageous due to their strong temporal coherence and ability to capture long-range dependencies in sequential data. However, they often suffer from error accumulation during sampling and can be computationally expensive for generating long sequences.  

More recently, diffusion-based audio models including Riffusion~\cite{forsgren2022riffusion}, AudioLDM~\cite{audioldm2}, DiffRhythm~\cite{diffrythm}, Möusai~\cite{mousai}, and Tango~\cite{tango} have enabled high-quality audio synthesis directly from text prompts. Diffusion models excel in producing realistic, high-fidelity audio and are more robust to error propagation compared to autoregressive methods. On the downside, they typically require lengthy iterative denoising steps, which makes inference slower and more resource-intensive.  

Recently, hybrid approaches that combine the strengths of both paradigms have emerged. Models such as Auffusion~\cite{Aufussion} and MagNet~\cite{magnet} integrate the fidelity and robustness of diffusion with the sequential modeling capacity of autoregressive transformers, offering a promising direction for efficient and controllable text-to-music generation.  

Control signals such as melody, chord progression, or rhythm have further improved conditioning and user controllability~\cite{ditto, musicontrolnet, cocomulla}. While these methods highlight the creative potential of large-scale generative models, they primarily focus on unconditional or text-conditioned generation, not editing.  

More recently, rectified flow (RF) has emerged as an alternative to classical diffusion for music generation~\cite{musflow, fluxplaysmusic} and editing~\cite{melodyflow}. By reformulating the denoising process into a continuous deterministic flow, RF enables faster and more stable text-to-music synthesis while preserving fine temporal and timbral details. This deterministic nature also makes RF particularly suitable for downstream tasks such as inversion and editing, laying the foundation for the approach we develop in this work.

\subsection{Music Editing}
Editing tasks for diffusion models are critical in practical music production but remain less explored compared to generation. Existing approaches typically follow two main directions. The first involves retraining or fine-tuning certain pretrained components of the model~\cite{adapter}. While effective in specific cases, these methods are limited because each type of edit requires a new round of fine-tuning. This process can be both computationally expensive and constrained by the scarcity of suitable training data.  

The second direction leverages pretrained generative models in a zero-shot fashion. For example, \emph{MusicMagus}~\cite{musicmagus} demonstrated zero-shot editing by manipulating the latent semantics of diffusion models. However, such methods often remain restricted to editing music generated by the model itself, with performance dropping significantly on real-world audio inputs. Moreover, many existing systems rely on complex and precise prompt engineering, which creates a barrier for non-expert users.  

\begin{figure*}[!t]
  \centering
  \includegraphics[width=\textwidth]{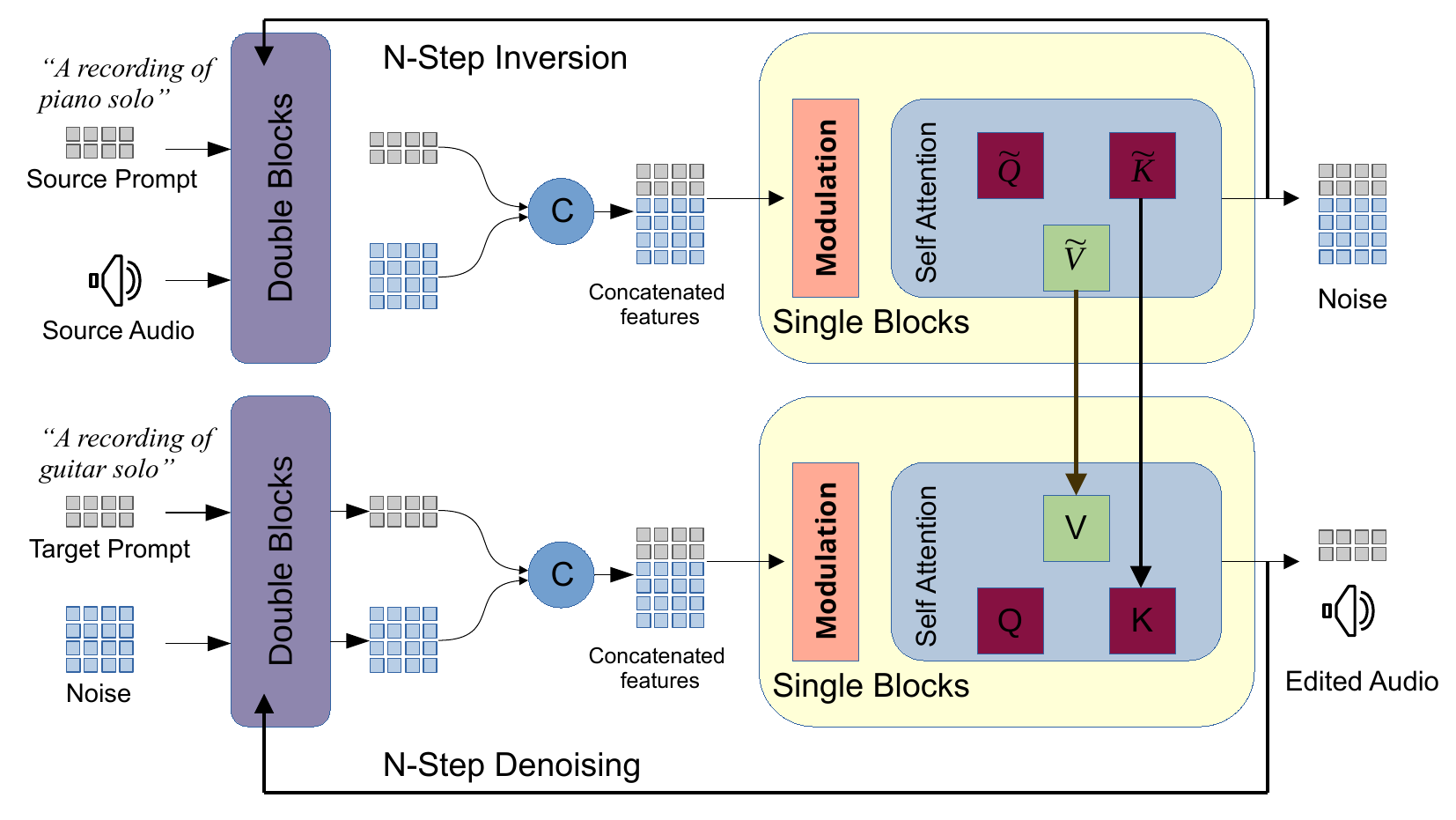}
  \caption{The source audio is first inverted into noise and then denoised to generate the edited audio. During denoising, the self-attention operations within the single blocks are modified according to their corresponding inversion steps. Note that the architecture comprises multiple single and double blocks, although only one of each is illustrated for clarity.}
  \label{fig:method}
\end{figure*}

In this work, we introduce \emph{MusRec}, a zero-shot framework built on pretrained rectified flow models for music editing. The overall process is illustrated in ~\ref{fig:method}. MusRec injects the self-attention features of the source music from a diffusion transformer directly into the editing process. Unlike prior approaches, it can operate effectively on real-world audio and generalizes across a wide range of editing tasks. Moreover, MusRec removes the need for prompt engineering during both reversal and editing, making music editing more accessible and practical.

\section{Background}

\subsection{Rectified Flow}

Let \(\pi_0\) and \(\pi_1\) denote two distributions on \(\mathbb{R}^d\) (in generative modeling \(\pi_0\) is a simple prior such as \(\mathcal{N}(0,I)\) and \(\pi_1\) is the data distribution). Rectified flow~\cite{liu2022flowstraightfastlearning} constructs intermediate states by coupling samples \((z_0,z_1)\) drawn from some joint coupling of \(\pi_0\) and \(\pi_1\) and then defining a \emph{linear} interpolation in time. Concretely, for \(t\in[0,1]\) we set
\begin{equation}
\label{eq:linear-interp}
z_t \;=\; \alpha_t\, z_0 \;+\; \beta_t\, z_1,
\end{equation}
where \(\alpha_t,\beta_t\) are scalar schedules satisfying \(\alpha_0=1,\beta_0=0\) and \(\alpha_1=0,\beta_1=1\). The canonical rectified-flow choice is \(\alpha_t=1-t,\ \beta_t=t\), yielding the straight path between \(z_0\) and \(z_1\).

Differentiating \eqref{eq:linear-interp} with respect to \(t\) gives the target instantaneous velocity along the interpolation:
\begin{equation}
\label{eq:target-velocity}
\dot z_t \;=\; \dot\alpha_t\, z_0 \;+\; \dot\beta_t\, z_1.
\end{equation}
Under the canonical schedule \(\alpha_t=1-t,\beta_t=t\), the RHS of \eqref{eq:target-velocity} is constant in \(t\),
\(\dot z_t = z_1 - z_0\), which is the defining ‘‘straight-line’’ property of rectified flow.

Rectified flow parameterizes a time-dependent velocity field \(v_\theta(z,t)\) (typically a neural network) and defines the generative dynamics by the ODE
\begin{equation}
\label{eq:rf-ode}
\frac{d z(t)}{dt} \;=\; v_\theta\bigl(z(t),t\bigr),
\end{equation}
with the goal that trajectories of \eqref{eq:rf-ode} when initialized from \(z_0\sim\pi_0\) transport mass to match \(\pi_1\). Training proceeds by \emph{velocity matching}: for \((z_0,z_1)\) drawn from the chosen coupling and \(t\sim\mathcal{U}[0,1]\), minimize the mean squared error between the network velocity and the target derivative,
\begin{equation}
\label{eq:velocity-loss}
\mathcal{L}(\theta)
\;=\;
\mathbb{E}_{(z_0,z_1),\,t}\!\bigl[\; \big\| v_\theta(z_t,t) \;-\; \dot z_t \big\|^2 \;\bigr],
\end{equation}
where \(z_t\) is given by \eqref{eq:linear-interp} and \(\dot z_t\) by \eqref{eq:target-velocity}. For the canonical linear schedule \(\dot z_t = z_1 - z_0\) and \eqref{eq:velocity-loss} reduces to matching the network output to the constant straight-line velocity.

In practice sampling (and inversion) integrate the learned ODE \eqref{eq:rf-ode} numerically. A standard explicit Euler discretization on a partition \(0=t_0<t_1<\dots<t_K=1\) yields the familiar update
\begin{equation}
\label{eq:euler-step}
z_{k+1} \;=\; z_k \;+\; (t_{k+1}-t_k)\, v_\theta(z_k,t_k),
\end{equation}
and higher-order integrators may be used in place of \eqref{eq:euler-step} to reduce discretization error. The rectified-flow design aims to make trajectories as straight (low-curvature) as possible so that coarse discretizations (small \(K\)) suffice for high-quality sampling; nevertheless, numerical integration error accumulates across steps and is the principal source of inversion/reconstruction error that RF-Solver later targets. 

\subsection{Inversion}

The goal of inversion is to recover the latent representation of observed data—such as images or audio—by reversing the generative dynamics. In diffusion models, one of the earliest and most widely adopted techniques is \textit{DDIM inversion}~\cite{songscore, ddim}. This method reconstructs the latent by progressively injecting noise predicted by the model at each forward step. Although this strategy succeeds in producing approximate reconstructions, it is inherently sensitive to discretization, since numerical integration introduces cumulative error across the trajectory. As a result, the final recovered signal may diverge from the original input. To mitigate this issue, several works~\cite{nti, stsl, elarabawy2022direct, negprompt} have explored improved inversion procedures. These approaches differ in implementation, yet all remain constrained by the underlying assumptions of diffusion-based dynamics.

In contrast, research on inversion within rectified flow models is still at an early stage. For instance, RF-Prior~\cite{yang2024text} applies score distillation to backtrack data into the latent domain, but the reliance on repeated optimization steps makes it computationally demanding. Another direction, proposed by~\cite{rout2024rfinversion}, augments the system with an additional vector field conditioned on the input, which provides improved reconstructions. Nevertheless, this approach does not fundamentally resolve the inaccuracies inherent in the rectified flow’s native vector field. Consequently, the effectiveness of current techniques remains limited when applied to downstream tasks that require both high-fidelity reconstruction and stable editing.

RF-Solver~\cite{tamingrectifiedflow} addresses this issue from a different angle by directly reducing the numerical errors associated with the rectified flow vector field. Instead of modifying the conditioning strategy or relying on optimization-heavy procedures, RF-Solver reformulates the rectified flow ODE using a variation-of-constants decomposition. This separates the system into linear and nonlinear components, with the nonlinear residual approximated through a high-order Taylor expansion. Such a treatment provides a substantially more accurate approximation of the trajectory during both forward sampling and reverse inversion. Importantly, the method is training-free and thus applicable to any pretrained rectified flow model. In practice, RF-Solver significantly enhances inversion fidelity, yielding reconstructions that more closely preserve input details, while simultaneously improving editability and generation quality compared to existing solvers.

Beyond RF-Solver, a few recent studies have also explored alternative numerical schemes to further improve the inversion process in rectified flow models. FireFlow~\cite{fireflow} proposes a second-order integration approach that delivers accurate inversions with noticeably fewer function evaluations, striking a practical balance between computational efficiency and reconstruction quality. Similarly, ABM-Solver~\cite{abmsolver} adopts an Adams--Bashforth--Moulton predictor--corrector method with adaptive step sizing, which helps maintain stability and produces more consistent edits across different cases. Although these methods were developed independently, they share the same motivation of making rectified flow inversion more reliable and efficient. They represent promising directions for future research and are worth attention as potential complements to solver-based approaches like RF-Solver.

\subsection{FLUX that Plays Music}

Recent work has extended rectified flow models beyond vision and into the audio domain. The system \textit{FLUX that Plays Music}~\cite{fluxplaysmusic} adapts the FLUX rectified flow transformer to text-to-music generation by operating in a latent mel-spectrogram space, demonstrating the versatility of rectified flow architectures across modalities.

The framework first converts raw waveforms into mel-spectrograms, which are then compressed into a lower-dimensional latent space through a variational autoencoder (VAE). All generative operations occur in this latent domain. Textual conditioning is provided through pretrained encoders such as T5~\cite{chung2024scaling}, which produces embeddings that capture semantic content, and CLAP~\cite{clap}, which produces embeddings aligned with audio, capturing both semantic content and audio-relevant attributes from prompts. Within the transformer backbone, generation alternates between \textit{double-stream} and \textit{single-stream} processing. Double-stream blocks handle text embeddings and music latents in parallel, with cross-attention allowing textual instructions to influence musical structure. Single-stream blocks then merge the two modalities, concatenating token-level features so that text and audio information can interact more directly. In addition, coarse-level features—such as global prompt vectors or temporal embeddings—are injected via modulation mechanisms that rescale hidden states.

At inference time, sampling begins from Gaussian noise $m(0)$, which is transported forward under rectified flow dynamics to produce a latent $m(1)$. This latent is decoded into a mel-spectrogram by the VAE decoder and finally rendered into an audible waveform by a vocoder. Due to the straightened transport trajectories of rectified flow, FluxMusic requires fewer integration steps than comparable diffusion-based text-to-audio models, thereby achieving faster generation.

Despite these advantages, the system is not without limitations. In practice, the generated music does not always faithfully reflect the input prompt: in particular, genres or styles that are underrepresented in the training distribution often lead to outputs that diverge from the intended semantics. This limitation arises from the generative model itself, which struggles to generalize to musical contexts it has not been adequately trained on. Ensuring robust prompt alignment therefore remains a significant challenge in text-to-music generation with rectified flow models.

\section{Methodology}

Our approach enables controlled editing of audio through a rectified flow–based generative framework. The process begins by encoding the source audio into a latent representation, which is then inverted into noise. During the subsequent denoising stage, the model reconstructs and edits the audio by modifying the self-attention operations according to the corresponding inversion steps. The editing is guided by a new text prompt, allowing semantic transformation of the source content while preserving its rhythmic and structural characteristics. An overview of the entire process is illustrated in Figure~\ref{fig:method}.

\subsection{Encoder}
\label{subsec:encoder}

We adopt the pretrained variational autoencoder (VAE) from AudioLDM2~\cite{liu2023audioldm} as our audio encoder. 
Given an input waveform, it is first converted into a mel-spectrogram representation through a TacotronSTFT-based frontend, which captures both spectral and temporal structure. 
The resulting features are then encoded into a compact latent representation by the VAE, effectively compressing high-dimensional audio information into a semantically meaningful latent space. 
This latent space serves as the foundation for downstream generative and editing tasks, enabling the model to operate on a continuous, information-rich representation of sound.

\subsection{RF-Solver}
\label{subsec:rfsolver}

The standard Rectified Flow (RF) sampler exhibits strong generative performance but struggles with inversion and reconstruction tasks due to cumulative errors at each timestep. These errors originate from the approximate solution of the rectified flow ordinary differential equation (ODE), which in prior work is estimated using a first-order Euler discretization~\cite{liu2022rectified}. To address this limitation, the RF-Solver method~\cite{tamingrectifiedflow} introduces a higher-order numerical scheme that provides a more accurate ODE approximation.

Starting from the continuous form of the rectified flow,
\begin{equation}
    \frac{d\boldsymbol{Z}_t}{dt} = \boldsymbol{v}_\theta(\boldsymbol{Z}_t, t),
\end{equation}
the method applies a Taylor expansion of $\boldsymbol{v}_\theta(\boldsymbol{Z}_\tau, \tau)$ around timestep $t_i$ and integrates it analytically, leading to the following $n$-th order approximation:
\begin{equation}
    \boldsymbol{Z}_{t_{i-1}} = \boldsymbol{Z}_{t_i} 
    + \sum_{k=0}^{n-1} \frac{(t_{i-1}-t_i)^{k+1}}{(k+1)!} 
    \boldsymbol{v}_\theta^{(k)}(\boldsymbol{Z}_{t_i}, t_i)
    + \mathcal{O}(h_i^{n+1}),
\end{equation}
where $\boldsymbol{v}_\theta^{(k)}$ denotes the $k$-th order time derivative of $\boldsymbol{v}_\theta$, and $h_i = t_{i-1} - t_i$.

In practice, the authors find that a second-order approximation ($n=2$) effectively mitigates reconstruction errors. The resulting update rule, termed RF-Solver, is:
\begin{align}
\label{equ:rfsolver}
\boldsymbol{Z}_{t_{i-1}} &= \boldsymbol{Z}_{t_i} 
    + (t_{i-1}-t_i)\,\boldsymbol{v}_\theta(\boldsymbol{Z}_{t_i}, t_i) \notag \\
    &\quad + \tfrac{1}{2}(t_{i-1}-t_i)^2\,\boldsymbol{v}_\theta^{(1)}(\boldsymbol{Z}_{t_i}, t_i).
\end{align}

Since $\boldsymbol{v}_\theta^{(1)}$ cannot be derived analytically, it is estimated numerically via finite differences:
\begin{equation}
    \boldsymbol{v}_\theta^{(1)}(\boldsymbol{Z}_{t_i}, t_i)
    = \frac{\boldsymbol{v}_\theta(\boldsymbol{Z}_{t_i+\Delta t}, t_i+\Delta t)
    - \boldsymbol{v}_\theta(\boldsymbol{Z}_{t_i}, t_i)}{\Delta t},
\end{equation}
where $\Delta t$ is a small perturbation (set to $0.01$ in practice). 

This second-order solver substantially reduces the local ODE error from $\mathcal{O}(h_i^2)$ to $\mathcal{O}(h_i^3)$, enabling more accurate inversion and reconstruction.

\subsection{Attention Feature Replacement Strategies}
\vspace{-0.1cm}
\setlength{\abovedisplayskip}{6pt}
\setlength{\belowdisplayskip}{6pt}

To explore the effect of feature-level guidance during the denoising process, several attention feature replacement strategies within the diffusion transformer architecture were further experimented with. The goal of these experiments is to investigate how reusing intermediate representations from the inversion stage can improve controllability and structure preservation in the generated outputs. Inspired by prior work on attention-level feature reuse in rectified flow models~\cite{tamingrectifiedflow}, three approaches were designed to modify the self-attention operation of the velocity prediction network $\boldsymbol{v}_\theta$ during the denoising process. In this setup, the focus is exclusively on the single transformer blocks, as they integrate information from both the source content and the conditioning input through unified modulation. While the double blocks in the underlying architecture process text and audio features separately, the single blocks concatenate these modalities, making them more suitable for controlled feature sharing. This design choice enables the model to leverage joint representations effectively, thereby enhancing its ability to preserve the structural and semantic characteristics of the source sample during generation.

During inversion, we cache the intermediate key and value tensors, $\{\mathcal{\widetilde{K}}_{t_k}^m\}$ and $\{\mathcal{\widetilde{V}}_{t_k}^m\}$, from the self-attention modules in the last $M$ transformer blocks across the final $n$ timesteps:
\begin{equation}
    \boldsymbol{\widetilde{F}}^m_{t_k} = \text{Attention}(\mathcal{\widetilde{Q}}_{t_k}^m, \mathcal{\widetilde{K}}_{t_k}^m, \mathcal{\widetilde{V}}_{t_k}^m),
\end{equation}
where $m \in \{1, \dots, M\}$ indexes the transformer blocks and $k \in \{N-n, \dots, N\}$ denotes the inversion timesteps. These features encode localized semantic and structural cues from the source sample.

In the denoising phase, we replace the standard self-attention mechanism
\begin{equation}
    \boldsymbol{F}_{t_k}^m = \text{Attention}(\mathcal{Q}_{t_k}^m, \mathcal{K}_{t_k}^m, \mathcal{V}_{t_k}^m)
\end{equation}
with modified formulations that inject the cached inversion features according to three strategies:
\begin{itemize}
    \item \textbf{(1) Value Replacement:} Replace only the value tensor with its cached counterpart,
    \begin{equation}
        \boldsymbol{F}_{t_k}^{m\prime} = \text{Attention}(\mathcal{Q}_{t_k}^m, \mathcal{K}_{t_k}^m, \mathcal{\widetilde{V}}_{t_k}^m),
    \end{equation}
    allowing the denoising process to reuse localized feature representations while maintaining the original attention distribution~\cite{crossattentioncontrol, plugandplay}.
    
    \item \textbf{(2) Key Replacement:} Replace only the key tensor with the cached key,
    \begin{equation}
        \boldsymbol{F}_{t_k}^{m\prime} = \text{Attention}(\mathcal{Q}_{t_k}^m, \mathcal{\widetilde{K}}_{t_k}^m, \mathcal{V}_{t_k}^m),
    \end{equation}
    emphasizing structural correspondence between the inversion and denoising phases~\cite{manifoldrepresentationkeyvision}.
    
    \item \textbf{(3) Key–Value Replacement:} Replace both the key and value tensors simultaneously,
    \begin{equation}
        \boldsymbol{F}_{t_k}^{m\prime} = \text{Attention}(\mathcal{Q}_{t_k}^m, \mathcal{\widetilde{K}}_{t_k}^m, \mathcal{\widetilde{V}}_{t_k}^m),
    \end{equation}
    effectively aligning both the attention map and the feature content with the inversion trajectory.
\end{itemize}

Through these experiments, we analyze how different forms of attention-level feature reuse influence reconstruction fidelity, edit consistency, and semantic controllability in generative tasks. This investigation provides insight into the role of cross-attention dynamics in rectified flow–based generation and their applicability to complex modalities such as music and audio.

\subsection{Classifier-Free Guidance}
\vspace{-0.1cm}
\setlength{\abovedisplayskip}{6pt}
In rectified flow models, classifier-free guidance (CFG) is applied by interpolating between conditional and unconditional velocity fields to modulate the strength of conditioning. Formally, given the conditional velocity field $v_\theta(x_t, y, t)$ and the unconditional one $v_\theta(x_t, \varnothing, t)$, the guided velocity $\hat{v}_\theta$ is defined as:
\[
\hat{v}_\theta(x_t, y, t) = v_\theta(x_t, \varnothing, t) + s \big(v_\theta(x_t, y, t) - v_\theta(x_t, \varnothing, t)\big),
\]
where $s$ denotes the CFG scale controlling the influence of the conditioning signal $y$. Higher $s$ values amplify the semantic conditioning, whereas lower values prioritize fidelity to the source.
In the experiments, classifier-free guidance (CFG) was employed to control the conditioning strength during both the inversion and denoising processes. The base model, \textit{FLUX that Plays Music}~\cite{fluxplaysmusic}, was trained with a fixed negative prompt of \textit{"low quality, gentle"}. Consequently, the same negative prompt was adopted across all experiments to maintain consistency with the model’s training distribution. Attempts to introduce alternative negative prompts resulted in degraded performance. For instance, during timbre transfer, \textit{"A recording of target instrument"} and \textit{"A recording of source instrument"} were used as positive and negative prompts during the denoising~\cite{adapter}. However, this configuration failed to yield meaningful results, likely due to the model’s reliance on its original negative prompt during training.

For the inversion process, the CFG scale was set to $1$, whereas the model’s default value during generation is $7$. Using a high CFG value (e.g., $7$) in inversion was observed to push latent representations into regions of the latent space that are difficult to guide back to meaningful and attribute-aligned states, thereby hindering effective editing while preserving melody and rhythm. Conversely, during the denoising stage, the CFG scale was increased to $20$ to strongly emphasize the new conditioning prompt, ensuring that the model incorporated the desired semantic changes while maintaining musical coherence.

\begin{figure*}[!t]
  \centering
  \includegraphics[width=\textwidth]{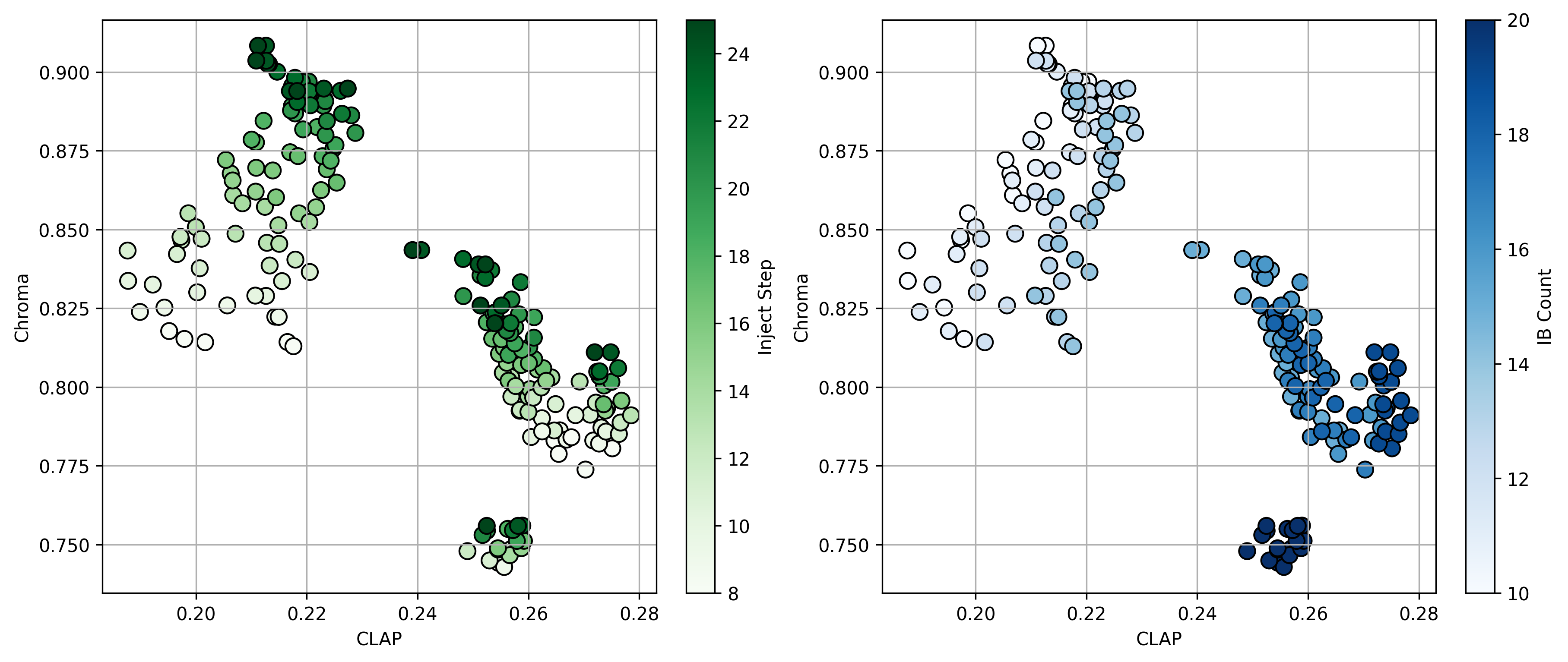}
   \caption{Transferability--fidelity trade-off effects of injection steps and IB (injection block) count on the timbre transfer task. The diagram shows the results of injecting the value (\(V\)) components of the attention mechanism into generation, i.e., how \(V\)-injection affects fidelity and transferability of the edited audio. For results of injecting the key (\(K\)) components or both key and value (\(K+V\)), and for all related results of genre transfer, see the Appendix.}
  \label{fig:comparison}
\end{figure*}

\section{Experiments}
\subsection{Datasets}
For our experiments, we curated two small yet high-quality datasets, each comprising 40 music clips collected from publicly available sources on YouTube, one designed for genre transfer and the other for timbre transfer. Each audio sample was manually selected to ensure clear instrument or genre distinction and minimal background noise. The clips were resampled to a sampling rate of 16 kHz and trimmed or segmented to a uniform duration of 10 seconds.  

The timbre dataset covers a diverse set of instrumental categories, including \textit{electric guitar}, \textit{flute}, \textit{piano}, \textit{violin}, and \textit{acoustic guitar}. These instruments were chosen to provide a balanced range of harmonic, percussive, and timbral characteristics, enabling a comprehensive evaluation of the model’s editing and timbre transfer capabilities. The genre dataset, on the other hand, encompasses a variety of musical styles, including \textit{pop}, \textit{jazz}, \textit{rock}, and \textit{hip-hop}, to assess the model’s effectiveness in capturing and transferring stylistic attributes across distinct genre domains.

\subsection{Baselines}
\label{subsec:baselines}

To validate the effectiveness of our method, we compare against four strong prior models widely used for text-to-music tasks.

\begin{itemize}
    \item \textbf{AudioLDM2 }:
\textit{AudioLDM2} \cite{audioldm2} is a latent diffusion model for text-to-audio (including music and sound effects), which conditions on text embeddings produced by CLAP and Flan-T5 and uses a U-Net-style architecture with cross-attention conditioning. For editing, we follow an SDEdit-style strategy~\cite{meng2021sdedit}: we partially apply the forward diffusion (i.e., adding noise) to the input audio up to a timestep $t_{\text{edit}}$, where $t_{\text{edit}} < T$ represents the noise level from which the reverse denoising process is initiated. The model then performs the reverse diffusion conditioned on the editing prompt to generate the edited audio.

    \item \textbf{MusicGen}:  
As a contrasting baseline, we employ \textit{MusicGen} \cite{musicgenstem}. MusicGen is a Transformer-based text-to-audio model that generates discrete audio tokens rather than diffusion in latent space. In particular, we use the \textit{MusicGen-Melody (1.5B)} variant, which allows conditioning on melody via a chromagram proxy.  
In our setup, we feed the edit prompt as text ($y$) and condition on the chromagram of the source audio ($x$), letting MusicGen generate the modified audio $\tilde{x}$ under this combined conditioning.

    \item \textbf{ZETA (DDPM Inversion):}  
\textit{Zero-Shot Unsupervised and Text-Based Audio Editing Using DDPM Inversion}~\cite{zeta} introduces two complementary modes: \textit{ZETA} and \textit{ZEUS}. ZETA performs text-guided editing by inverting the diffusion process for an input audio $x$ and steering the denoising trajectory using a textual prompt $y$. We include ZETA as a baseline to evaluate text-based audio editing performance under the DDPM inversion framework.
  
    \item \textbf{FluxMusic:}  
\textit{FluxMusic}~\cite{fluxplaysmusic} is a rectified flow transformer model for text-to-music generation. In our setup, we use the RF-solver~\cite{tamingrectifiedflow} to invert the input audio $x$ into its latent representation and then generate the edited music $\tilde{x}$ under text conditioning $y$. This approach enables semantically guided audio editing while maintaining the structural coherence of the original piece.

\end{itemize}
In addition, we considered several recent models, including \textit{MusicMagus}~\cite{musicmagus} and \textit{TransPlayer}~\cite{transplayer}, but did not include them in our direct comparison. \textit{MusicMagus} showed limited performance on our editing tasks due to its limitations on real music data and was therefore not included in the quantitative results table, while \textit{TransPlayer} supports only limited edit tasks, which diverges from the objective of our study. We also examined \textit{MelodyFlow}~\cite{melodyflow} and \textit{SteerMusic}~\cite{steermusic}; however, at the time of writing this paper, neither model had publicly available code or checkpoints, preventing a fair evaluation. Finally, the \textit{Audio Prompt Adapter}~\cite{adapter} was excluded, as its checkpoints were recently removed.

\subsection{Objective Metrics}
\label{subsec:metrics}

We evaluate our method using two complementary objective metrics that assess both transferability and fidelity:  
\begin{itemize}
     
    \item \textbf{CLAP Similarity:} CLAP~\cite{clap} evaluates the semantic alignment between audio and text by mapping both modalities into a shared embedding space through contrastive learning. The cosine similarity between the CLAP embeddings of $\tilde{x}$ and the conditioning text $y$ measures how well the generated audio reflects the intended semantic meaning.

    \item \textbf{Chroma Similarity:} To evaluate the fidelity of the generated audio, we compute the chroma similarity between the original audio $x$ and the edited audio $\tilde{x}$. This metric captures harmonic and rhythmic correspondence by comparing their chromagrams, extracted using the Constant-Q Transform (CQT) chroma method implemented in \texttt{librosa}~\cite{mcfee2015librosa}. Framewise cosine similarity between the chroma features provides a quantitative measure of how well the edited sample preserves the musical structure of the source.  

    \item \textbf{CQT-1 PCC:} The Constant-Q Transform (CQT)~\cite{brown1991cqt} represents audio on a logarithmic frequency scale, reflecting human pitch perception. We compute the Pearson Correlation Coefficient (PCC) between the CQT magnitude spectra of the original audio $x$ and the edited audio $\tilde{x}$. Higher values indicate stronger spectral correspondence, suggesting that harmonic and timbral structures are well preserved.

    \item \textbf{Fréchet Audio Distance (FAD):} To assess the perceptual quality and distributional similarity of the generated audio, we compute the Fréchet Audio Distance (FAD) between the real and generated samples. Analogous to the Fréchet Inception Distance (FID) used in image generation, FAD measures the distance between two multivariate Gaussian distributions fitted to embeddings extracted from a pretrained audio feature extractor (VGGish). These embeddings capture high-level perceptual attributes such as timbre, texture, and overall audio quality. A lower FAD score indicates that the generated audio is closer in distribution to real audio, reflecting higher perceptual realism and better generative performance.

\end{itemize}

\subsection{Subjective Metrics}
\label{subsec:subjective_metrics}

To complement the objective evaluations, we conduct a subjective assessment of the perceptual and semantic quality of the generated music. Following the ITU-T recommendations for subjective evaluation of multimedia content~\cite{itut1999video,itut1996audio}, we employ two Mean Opinion Score (MOS) metrics: \textit{MOS-T} and \textit{MOS-P}.

\textbf{MOS-T} measures the perceived alignment between the generated audio and its corresponding target prompt. Participants rate, on a 5-point Likert scale, how well the musical content reflects the semantics, emotion, and style expressed in the text prompt.

\textbf{MOS-P} evaluates how well the edited audio $\tilde{x}$ preserves the perceptual characteristics of the source audio $x$, including timbre, rhythm, and overall musical structure. Higher MOS-P values indicate that the edited output maintains greater perceptual similarity to the original recording while integrating the intended edits naturally.

\begin{table*}[t]
\centering
\caption{The objective evaluation results on the timbre transfer.}
\begin{tabular}{lcccccc}
\hline
\textbf{Model} & \textbf{Type} & \textbf{CLAP $\uparrow$} & \textbf{Chroma $\uparrow$} & \textbf{CLAP+Chroma Avg. $\uparrow$} & \textbf{CQT-1 PCC $\uparrow$} & \textbf{FAD $\downarrow$} \\
\hline
MusicGen & Supervised  & .220 & .757 & .489 & .274 & 5.320 \\
AudioLDM2 & Zero-shot  & .235 & .820 & .527  & .563 & \textbf{3.574} \\
Zeta & Zero-shot  & .224 & .813 & .518 & .560 & 5.693 \\
FluxMusic & Zero-shot  & .220 & .756  & .488 & .464 & 5.403 \\
\hline
MusRec K Injection\textbf{(ours)} & Zero-shot  & \textbf{.262} & .718 & .490 & .366 & 7.018 \\
MusRec KV Injection\textbf{(ours)} & Zero-shot  & \underline{.237} & \textbf{.851} & \textbf{.543} & \textbf{.600} & \underline{4.265} \\
MusRec V Injection\textbf{(ours)} & Zero-shot  & .236 & \underline{.843} & \underline{.535} & \underline{.583} & 4.605 \\
\hline
\end{tabular}
\label{tab:objective_timbre}
\end{table*}

\begin{table*}[t]
\centering
\caption{The objective evaluation results on the genre transfer.}
\begin{tabular}{lcccccc}
\hline
\textbf{Model} & \textbf{Type} & \textbf{CLAP $\uparrow$} & \textbf{Chroma $\uparrow$} & \textbf{CLAP+Chroma Avg. $\uparrow$} & \textbf{CQT-1 PCC $\uparrow$} & \textbf{FAD $\downarrow$} \\
\hline
MusicGen & Supervised  & .454 & .754 & .604 & .129 & 9.790 \\
AudioLDM2 & Zero-shot  & \textbf{.585} & .698 & .641 & .153 & 8.782 \\
Zeta & Zero-shot & .531 & .762 & .646 & .315 & 7.158 \\
FluxMusic & Zero-shot & .524 & .771 & .647 & .354 & 6.774 \\
\hline
MusRec K Injection\textbf{(ours)} & Zero-shot & \underline{.547} & .754 & .650 & .225 & 11.398 \\
MusRec KV Injection\textbf{(ours)} & Zero-shot & .545 & \underline{.797} & \textbf{.671} &  \underline{.424} & \textbf{5.433} \\
MusRec V Injection\textbf{(ours)} & Zero-shot & .537 & \textbf{.799} & \underline{.668} & \textbf{.433} & \underline{5.662} 
\\
\hline
\end{tabular}
\label{tab:objective_genre}
\end{table*}

\section{Results and Discussion}

\subsection{Hyperparameter Selection}
We discovered that several hyperparameters significantly affect the quality of edited music. To systematically study their impact, we conducted experiments analyzing the effects of different parameters. In our setup, five hyperparameters can be tuned depending on the task and the source audio: the number of diffusion steps, the target classifier-free guidance scale, the source classifier-free guidance scale, the number of injection steps, and the injection block count (IB count).

Since jointly optimizing all five parameters would result in an impractically large search space, we focused our detailed analysis on the injection steps and IB count, while determining suitable values for target CFG, source CFG and number of steps empirically. We observed that target CFG values in the range of 15--25 generally yield the best performance across tasks, while a value of 1 for the source CFG provides stable results. Similarly, we set the number of diffusion steps to 25. Although increasing the number of steps improves performance, we chose 25 to maintain a balance between quality and computational efficiency. While the base model, FluxMusic, generates music using a default of 50 diffusion timesteps, we reduced this number to 25 to accelerate the generation process and lower computational cost.

\textbf{Injection steps} determine at which diffusion steps the model injects the attention mechanism information derived from the corresponding inversion steps (as illustrated in Figure~\ref{fig:comparison}). Increasing the number of injection steps enhances fidelity but reduces transferability. For instance, setting the injection step too high causes the model to preserve excessive acoustic details from the input, leading to edited outputs that sound similar to the original audio but may not accurately reflect the editing command.

\textbf{IB count} specifies after which single block within each injection step the attention injection occurs. For example, if there are \( n \) single blocks and the IB count is \( m \) (where \( m < n \)), the injection happens after the \( m \)-th block. As shown in Figure~\ref{fig:comparison}, increasing the IB count improves transferability but decreases fidelity, indicating a trade-off between these two factors.

\subsection{Objective Results}
\label{subsec:objective_evaluation}

We conduct an objective evaluation to quantitatively assess the performance of the proposed MusRec model on both timbre and genre transfer tasks. The evaluation relies on four key metrics: CLAP similarity, which measures semantic alignment between the generated and target audio; Chroma similarity, which reflects harmonic fidelity to the source; CQT-1 PCC, which captures spectral correlation and timbral preservation between the source and generated audio; and Fréchet Audio Distance (FAD), which estimates perceptual realism by comparing feature distributions of generated and real audio.

Table~\ref{tab:objective_timbre} presents the results for the timbre transfer task. Among all models, MusRec K Injection attains the highest CLAP similarity score, demonstrating the strongest semantic alignment with the conditioning prompt, followed closely by MusRec KV Injection. In terms of harmonic and spectral fidelity, as measured by Chroma similarity and CQT-1 PCC, MusRec KV Injection achieves the best performance, with MusRec V Injection ranking second in both metrics. When considering the average of Chroma and CLAP similarity, MusRec KV Injection again provides the most balanced outcome, indicating effective integration of semantic and acoustic cues. Regarding perceptual realism, assessed via FAD, AudioLDM2 yields the lowest score, while MusRec KV Injection ranks second, confirming that it maintains high perceptual quality while preserving fidelity to the source.

Table~\ref{tab:objective_genre} presents the results for the genre transfer task. The overall trends are consistent with the timbre transfer evaluation. In terms of CLAP similarity, AudioLDM2 achieves the highest score, followed by MusRec K Injection, indicating strong semantic alignment with the target genre. For harmonic and spectral measures—Chroma similarity and CQT-1 PCC—MusRec V Injection performs best, with MusRec KV Injection ranking second, reflecting superior preservation of tonal and timbral characteristics. When averaging CLAP and Chroma scores, MusRec KV Injection attains the best overall balance between semantic consistency and harmonic fidelity, closely followed by MusRec V Injection. Regarding perceptual realism, as measured by FAD, MusRec KV Injection achieves the lowest score, with MusRec V Injection in second place, demonstrating high perceptual quality and effective genre adaptation.

Overall, the objective evaluation demonstrates that incorporating both key and value attention injections provides the most balanced performance across timbre and genre transfer tasks. The MusRec KV Injection variant consistently achieves strong trade-offs between semantic alignment, harmonic fidelity, and perceptual realism, while the V Injection configuration excels in preserving tonal characteristics and producing perceptually coherent outputs. In contrast, the K Injection variant favors semantic transfer, achieving higher CLAP alignment but with a modest reduction in spectral fidelity.

\begin{table*}[t]
\centering
\begin{minipage}{0.47\textwidth}
\centering
\captionof{table}{The subjective evaluation results on the timbre transfer.}
\label{tab:timbre_subjective}
\begin{tabular}{lccc}
\toprule
\textbf{Model} & \textbf{MOS-T $\uparrow$} & \textbf{MOS-P $\uparrow$} & \textbf{Overall $\uparrow$} \\
\midrule
AudioLDM2                 & 3.10 & 3.33 & 3.21 \\
MusicGen                  & 3.33 & 2.62 & 2.98 \\
Zeta                      & 3.57 & 3.57 & 3.57 \\
FluxMusic                 & 3.43 & 3.71 & 3.57 \\
\hline
MusRec KV Injection (\textbf{ours}) & \textbf{4.05} & \underline{4.14} & \textbf{4.10} \\
MusRec V Injection (\textbf{ours})  & \underline{3.90} & \textbf{4.24} & \underline{4.07} \\
MusRec K Injection (\textbf{ours})  & 3.43 & 3.05 & 3.24 \\
\bottomrule
\end{tabular}
\end{minipage}
\hfill
\begin{minipage}{0.47\textwidth}
\centering
\captionof{table}{The subjective evaluation results on the genre transfer.}
\label{tab:genre_subjective}
\begin{tabular}{lccc}
\toprule
\textbf{Model} & \textbf{MOS-T $\uparrow$} & \textbf{MOS-P $\uparrow$} & \textbf{Overall $\uparrow$} \\
\midrule
AudioLDM2                 & \textbf{3.14} & 1.86 & 2.50 \\
MusicGen                  & 2.67 & 2.57 & 2.62 \\
Zeta                      & 2.71 & 3.76 & 3.24 \\
FluxMusic                 & \underline{2.95} & 3.62 & 3.29 \\
\hline
MusRec KV Injection (\textbf{ours}) & \textbf{3.14} & \underline{4.14} & \underline{3.64} \\
MusRec V Injection (\textbf{ours})  & \textbf{3.14} & \textbf{4.19} & \textbf{3.67} \\
MusRec K Injection (\textbf{ours})  & \underline{2.95} & 3.19 & 3.07 \\
\bottomrule
\end{tabular}
\end{minipage}
\end{table*}

\subsection{Subjective Results}

To evaluate the perceptual quality of the generated audio, we conducted an online subjective listening test using Google Forms with 21 participants, comprising 11 professional musicians and 10 ordinary listeners without formal musical training. To ensure reliable subjective evaluation, participants were recruited voluntarily online; the slight imbalance between professional and ordinary listeners (11 vs.\ 10) does not affect the overall analysis, as results were averaged separately across both groups. Each participant was randomly assigned one sample for genre transfer and one for timbre transfer. For each sample, they provided two ratings on a five-point Likert scale: the Mean Opinion Score for Timbre (MOS-T), reflecting the naturalness and timbral realism of the output, and the Mean Opinion Score for Perceptual Quality (MOS-P), indicating the overall perceptual quality of the transferred audio. The summarized results are shown in Tables~\ref{tab:timbre_subjective} and~\ref{tab:genre_subjective}, while the detailed breakdown by professional and ordinary listeners is provided in the Appendix. These subjective evaluations closely follow the same trends observed in the objective metrics, further confirming the consistency and reliability of the proposed framework.

Table~\ref{tab:timbre_subjective} presents the results for the timbre transfer task. Among all models, MusRec KV Injection achieves the highest overall perceptual and timbral quality, demonstrating excellent preservation of tonal attributes and structural coherence. MusRec V Injection also performs strongly, producing smooth and natural-sounding edits with consistent fidelity to the input recording. In contrast, MusRec K Injection prioritizes prompt adherence but with slightly reduced perceptual naturalness. All three MusRec variants outperform the baseline models, highlighting the advantage of the proposed MusRec models in generating perceptually convincing timbre transformations.

Table~\ref{tab:genre_subjective} reports the subjective evaluation for the genre transfer task. Similar to the timbre transfer results, MusRec V Injection delivers the most perceptually coherent and musically natural outputs, while MusRec KV Injection achieves a strong balance between genre adaptation and fidelity to the original material. The MusRec K Injection variant again emphasizes semantic adherence at a modest cost in perceptual realism. Baseline models show comparatively weaker performance, particularly in perceptual quality, reflecting limited generalization to recordings. 

Overall, the subjective results reinforce the findings of the objective evaluation: integrating both key and value conditioning leads to a balanced trade-off between text alignment and perceptual quality, while value-only conditioning excels in producing smooth and natural musical outputs. These findings validate the effectiveness of the proposed approach in achieving high-quality, zero-shot text-driven music editing on real-world audio.

\section{Conclusion}

In conclusion, this work presents a novel zero-shot music editing framework based on rectified flow modeling. The proposed method effectively edits the source music toward a target text prompt while preserving essential musical attributes such as timbre, melody, rhythm, and overall structural coherence. To the best of our knowledge, this is the first zero-shot music editing approach built upon rectified flow, capable of operating directly on real-world music recordings. 

Although FluxMusic, the underlying base model, exhibits limited capability in faithfully following textual prompts and producing high-fidelity outputs, our results demonstrate that the proposed editing mechanism substantially improves controllability and consistency. We believe that applying this framework to future rectified-flow-based music generation models with stronger priors and higher audio quality could further enhance its performance and generalization in real-world editing scenarios.

\bibliographystyle{IEEEtran}
\bibliography{references}

\clearpage
\appendix
\section{Appendix}
\subsection{Additional Results on Attention Injection Variants}
\label{sec:appendix}

\begin{figure}[h!]
  \centering
  \includegraphics[width=\linewidth]{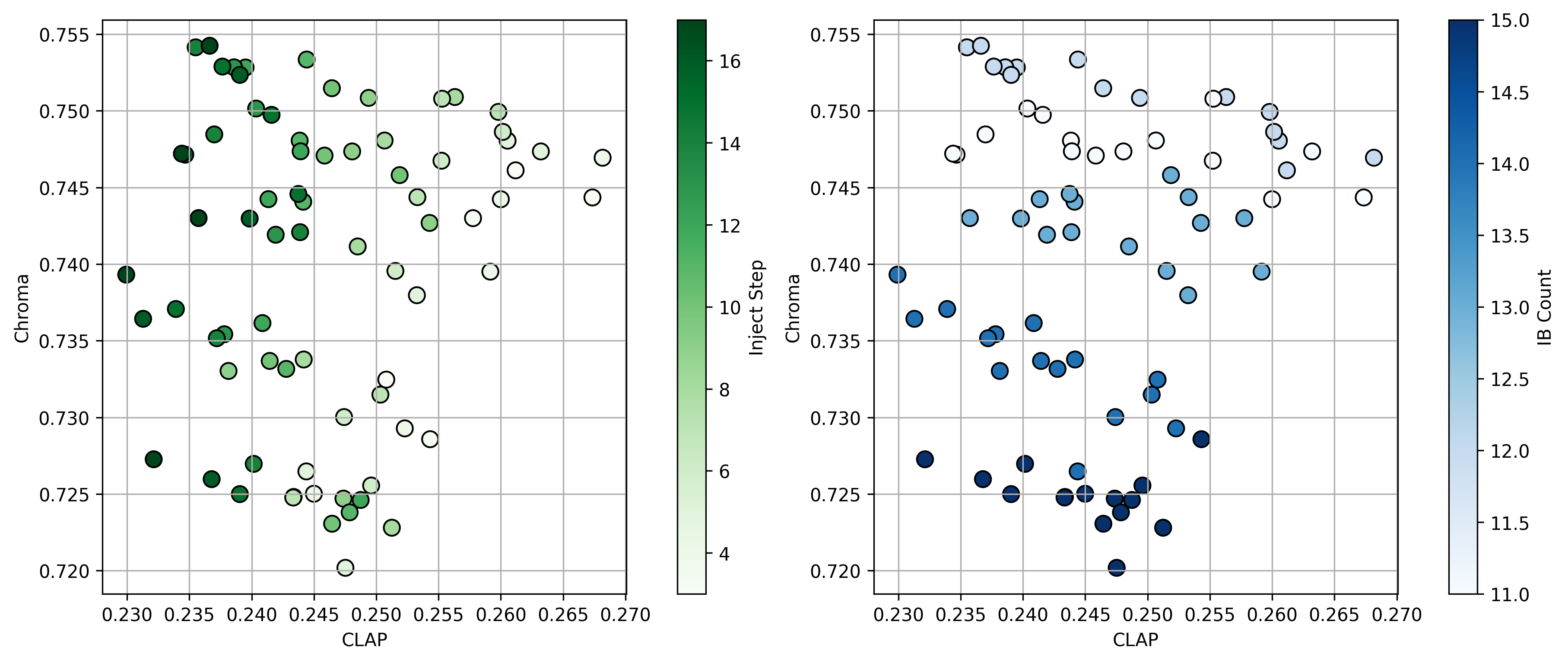}
  \caption{Results of injecting the key (\(K\)) components of the attention mechanism during timbre transfer task. Injecting \(K\) leads to moderate improvements in transferability but slightly weaker fidelity compared to \(V\)-injection, as less low-level acoustic information is preserved.}
  \label{fig:injection_K}
\end{figure}

\begin{figure}[h!]
  \centering
  \includegraphics[width=\linewidth]{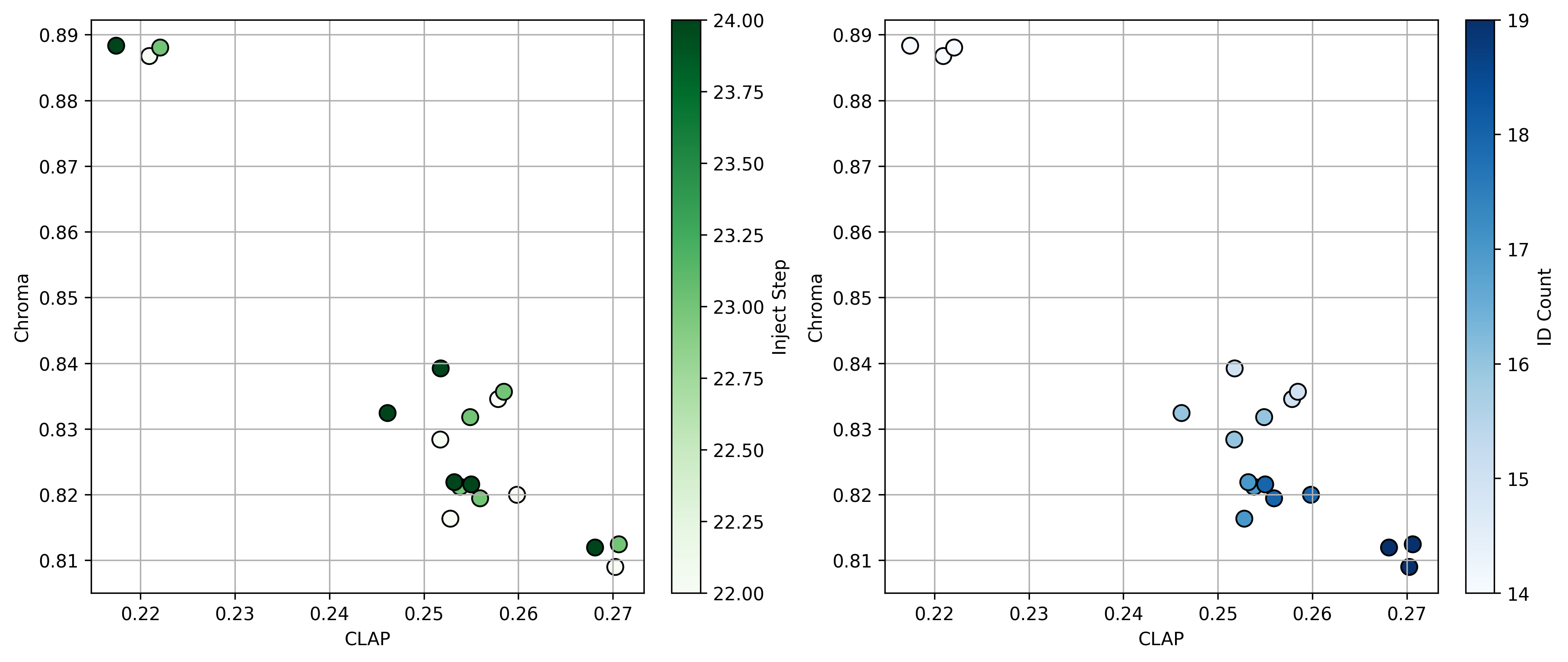}
  \caption{Results of injecting both key and value (\(K+V\)) components of the attention mechanism during timbre transfer task. Injecting \(K+V\) tends to balance fidelity and transferability, yielding more consistent timbre adaptation while retaining semantic control.}
  \label{fig:injection_KV}
\end{figure}

\begin{figure}[h!]
  \centering
  \includegraphics[width=\linewidth]{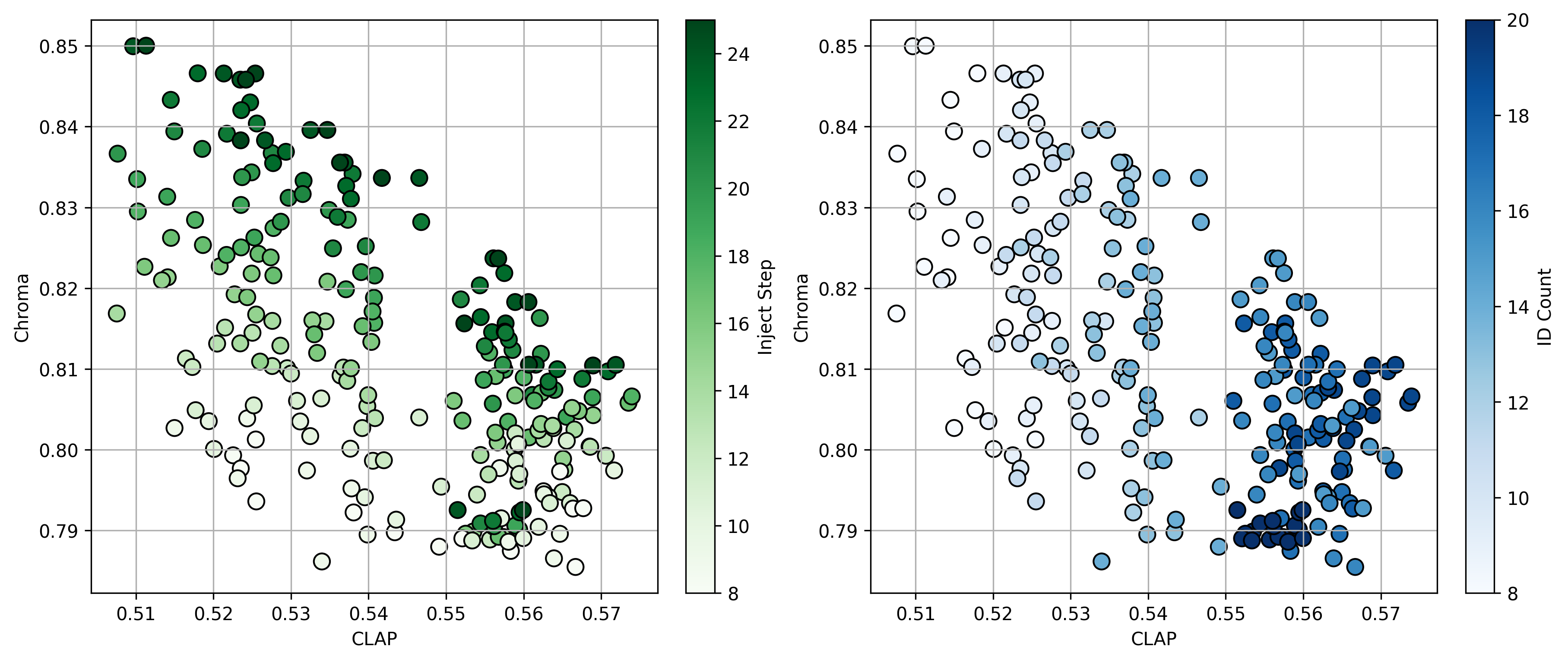}
  \caption{Results of injecting the value (V) components of the attention mechanism during genre transfer. Injecting V mainly preserves fidelity while limiting the degree of stylistic transfer, showing more stable tonal similarity across genres.}
  \label{fig:genre_injection_V}
\end{figure}

\begin{figure}[h!]
  \centering
  \includegraphics[width=\linewidth]{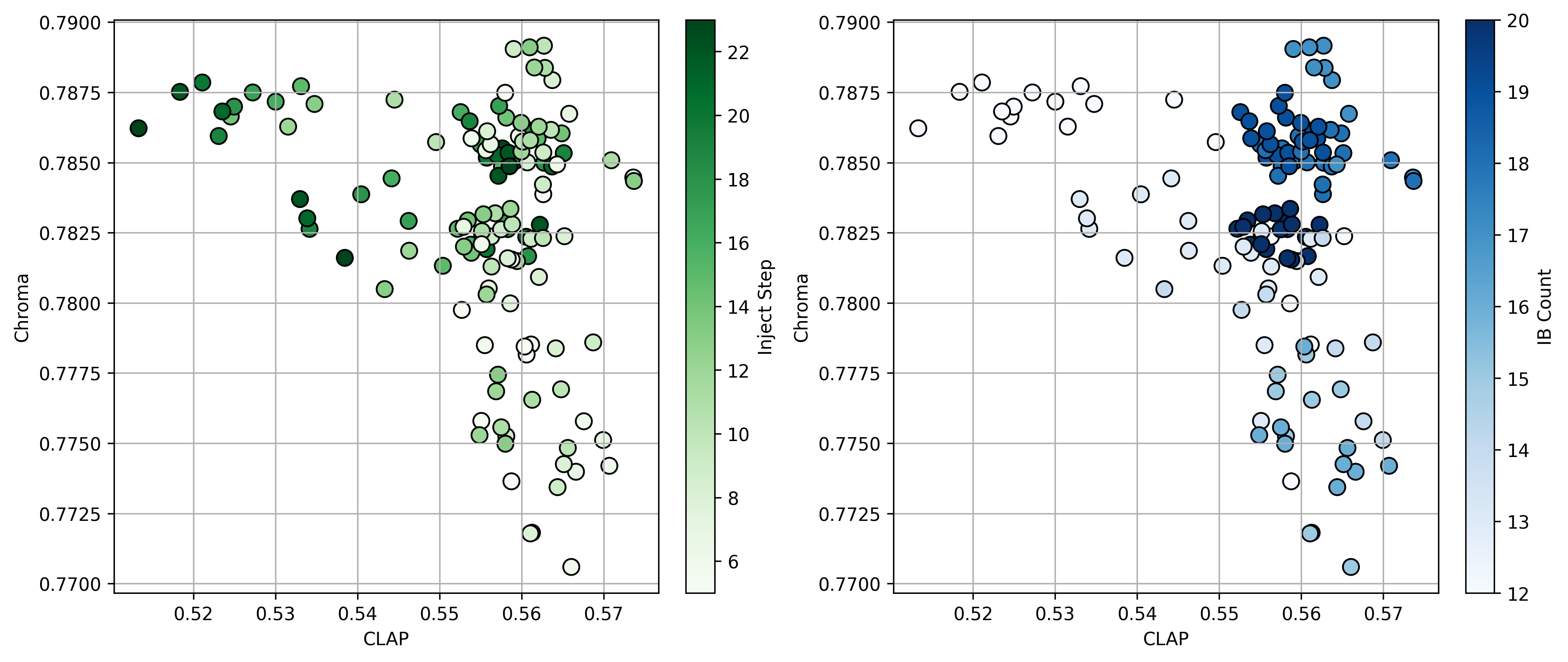}
  \caption{Results of injecting the key (K) components of the attention mechanism during genre transfer. Injecting only K emphasizes structural transferability but can reduce chroma fidelity, indicating that genre cues dominate over tonal preservation.}
  \label{fig:genre_injection_K}
\end{figure}

\begin{figure}[h!]
  \centering
  \includegraphics[width=\linewidth]{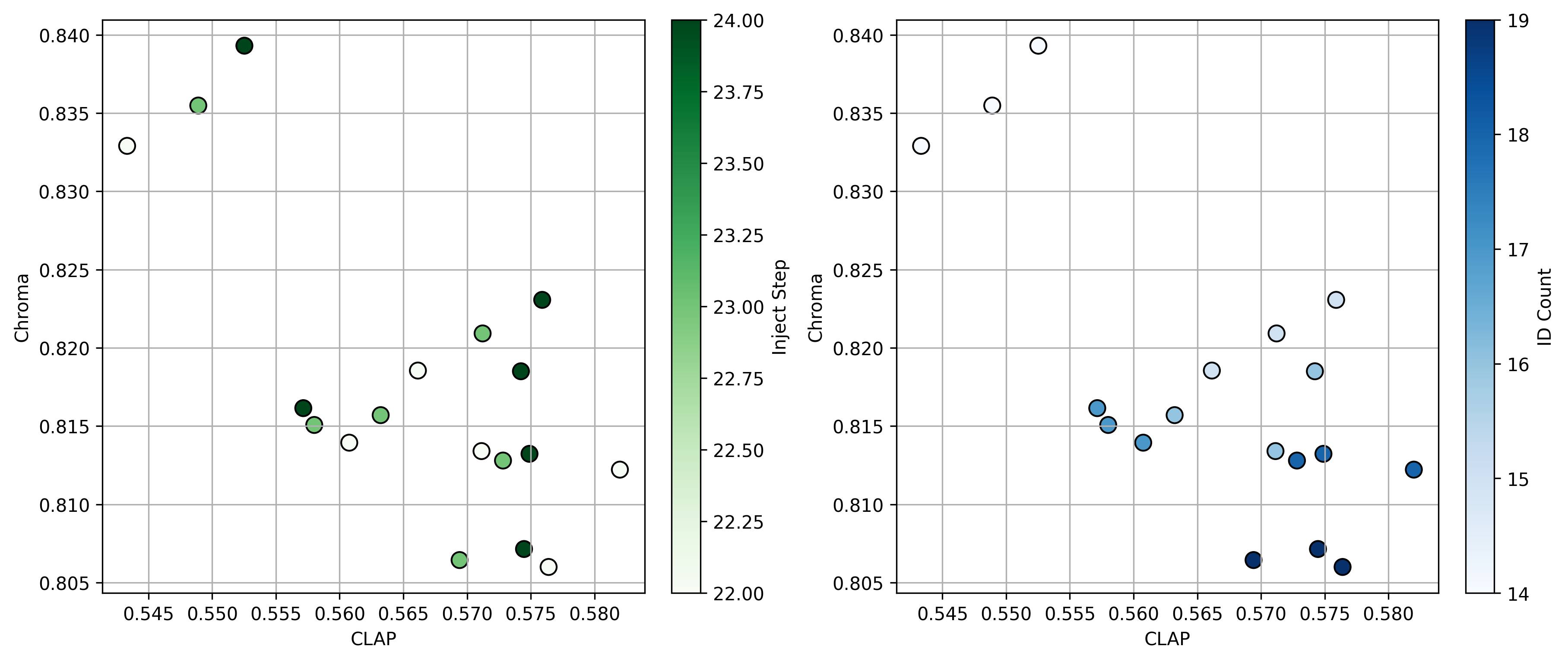}
  \caption{Results of injecting both key and value (K + V) components of the attention mechanism during genre transfer. Injecting K + V achieves a better balance between fidelity and transferability, enabling effective genre transformation while maintaining harmonic consistency.}
  \label{fig:genre_injection_KV}
\end{figure}

To complement the main results presented in Figure~\ref{fig:comparison}, we further investigate the effects of injecting the key (K) components and the combined key and value (K + V) components of the attention mechanism during the generation process. In addition to the timbre transfer experiments, we also include the results of injecting value (V), key (K), and key + value (K + V) components for the genre transfer task, as shown here. These extended analyses demonstrate that the choice of injected components directly influences the balance between fidelity and transferability. Specifically, V-injection preserves detailed timbral and genre-specific characteristics, K-injection promotes stronger adherence to the conditioning or editing command, and K + V-injection offers a balanced compromise between the two, achieving consistent transformations while maintaining perceptual and structural coherence.

\subsection{Full Subjective Results}
\begin{table*}[t]
\centering
\caption{The full subjective evaluation results on the timbre transfer.}
\begin{tabular}{lcccccc}
\toprule
 & \multicolumn{3}{c}{\textbf{MOS-T mean $\uparrow$}} & \multicolumn{3}{c}{\textbf{MOS-P mean $\uparrow$}} \\
\cmidrule(lr){2-4} \cmidrule(lr){5-7}
\textbf {Model} & \textbf{Overall} & \textbf{Professional Musicians} & \textbf{Ordinary Listeners} & \textbf{Overall} & \textbf{Professional Musicians} & \textbf{Ordinary Listeners} \\
\midrule
MusicGen & 3.33 & 3.55 & 3.10 & 2.62 & 2.55 & 2.70 \\
AudioLDM2 & 3.10 & 3.64 & 2.50 & 3.33 & 3.36 & 3.30 \\
Zeta & 3.57 & 3.55 & 3.60 & 3.57 & 3.55 & 3.60 \\
FluxMusic & 3.43 & 3.45 & 3.40 & 3.71 & \underline{3.82} & 3.60 \\
\hline
MusRec KV Injection (\textbf{ours}) & \textbf{4.05} & \textbf{4.27} & \underline{3.80} & \underline{4.14} & \textbf{4.27} & \underline{4.00} \\
MusRec V Injection (\textbf{ours}) & \underline{3.90} & \underline{3.82} & \textbf{4.00} & \textbf{4.24} & \textbf{4.27} & \textbf{4.20} \\
MusRec K Injection (\textbf{ours}) & 3.43 & 3.36 & 3.50 & 3.05 & 2.91 & 3.20 \\
\bottomrule
\end{tabular}
\label{tab:full_subjective_timbre}
\end{table*}

\begin{table*}[t]
\centering
\caption{The full subjective evaluation results on the genre transfer.}
\begin{tabular}{lcccccc}
\toprule
 & \multicolumn{3}{c}{\textbf{MOS-T mean $\uparrow$}} & \multicolumn{3}{c}{\textbf{MOS-P mean $\uparrow$}} \\
\cmidrule(lr){2-4} \cmidrule(lr){5-7}
\textbf {Model} & \textbf{Overall} & \textbf{Professional Musicians} & \textbf{Ordinary Listeners} & \textbf{Overall} & \textbf{Professional Musicians} & \textbf{Ordinary Listeners} \\
\midrule
MusicGen & 2.67 & \underline{2.91} & 2.40 & 2.57 & 2.45 & 2.70 \\
AudioLDM2 & \textbf{3.14} & \textbf{3.27} & 3.00 & 1.86 & 1.64 & 2.10 \\
Zeta & 2.71 & 2.45 & 3.00 & 3.76 & 4.18 & 3.30 \\
FluxMusic & \underline{2.95} & 2.73 & 3.20 & 3.62 & 3.73 & \underline{3.50} \\
\hline
MusRec KV Injection (\textbf{ours}) & \textbf{3.14} & \underline{2.91} & \textbf{3.40} & \underline{4.14} & \textbf{4.73} & \underline{3.50} \\
MusRec V Injection (\textbf{ours}) & \textbf{3.14} & \underline{2.91} & \textbf{3.40} & \textbf{4.19} & \underline{4.55} & \textbf{3.80} \\
MusRec K Injection (\textbf{ours}) & \underline{2.95} & 2.73 & \underline{3.20} & 3.19 & 3.27 & 3.10 \\
\bottomrule
\end{tabular}
\label{tab:full_subjective_genre}
\end{table*}

Tables~\ref{tab:full_subjective_timbre} and~\ref{tab:full_subjective_genre} present the full subjective evaluation results, including separate scores from professional musicians and ordinary listeners. The results reveal clear but complementary differences in perception between the two groups.

Professional musicians generally assigned higher MOS-T and MOS-P ratings for models that preserved timbral detail and musical structure, showing greater sensitivity to subtle artifacts or tonal imbalances. They consistently preferred \textit{MusRec KV Injection}, which provided the most faithful timbral transfer in both tasks, while \textit{MusRec V Injection} also received high ratings from participants. Among the baseline systems, \textit{FluxMusic} and \textit{Zeta} received comparatively better scores from professionals, while \textit{AudioLDM2} achieved a slightly higher MOS-T in genre transfer, suggesting that its tonal balance was appreciated despite its lower perceptual realism. In contrast, \textit{MusicGen} was rated the lowest among all models.

Ordinary listeners, on the other hand, tended to favor models that maintained overall musical coherence and recognizable style, even when minor distortions were present. For this group, \textit{MusRec V Injection} often received the highest perceptual ratings, as its outputs were smoother and easier to follow, while \textit{MusRec KV Injection} ranked slightly lower but remained among the top performers. Ordinary listeners also rated \textit{FluxMusic} and \textit{Zeta} relatively well among the baselines, whereas \textit{AudioLDM2} and \textit{MusicGen} were perceived as less consistent.

Across both listener groups, \textit{MusRec K Injection} was perceived as more prompt-aligned but slightly less natural, and the baseline models were consistently rated lower, particularly by professionals. Importantly, the relative ranking of models remained consistent across both groups, confirming that the improvements achieved by the proposed \textit{MusRec} variants are perceptually robust across varying levels of musical expertise.

\end{document}